\documentclass[aps,prd,twocolumn,superscriptaddress,showkeys,showpacs]{revtex4}
\usepackage{pdfsync}
\usepackage{mathptmx}
\usepackage[utf8]{inputenc}
\usepackage[T1]{fontenc}
\usepackage{slashed}
\usepackage{times}
\usepackage{amssymb,amsfonts,amsmath,amsthm}
\usepackage{dsfont,bbm}
\usepackage{dcolumn}
\usepackage{epsf}
\usepackage{graphicx}
\usepackage[caption=false]{subfig}
\usepackage{dsfont}
\usepackage{bm}
\usepackage{slashed}
\usepackage[active]{srcltx}
\usepackage[usenames]{color}

\begin{document}

\title{Inverse Radon transform and
the transverse-momentum dependent functions}

\author{I.~V.~Anikin}
\email{anikin@theor.jinr.ru}
\affiliation{Bogoliubov Laboratory of Theoretical Physics, JINR,
             141980 Dubna, Russia}
\author{L.~Szymanowski}
\email{Lech.Szymanowski@ncbj.gov.pl}
\affiliation{National Centre for Nuclear Research (NCBJ),
            02-093 Warsaw, Poland}

\begin{abstract}
We revisit the standard representation of the (inverse) Radon transform which is well-known in the mathematical literature.
We extend this representation to the case involving the parton distributions.
We have found the new additional contribution which is essentially
related to the generalized transverse-momentum dependent parton distribution and
double-distribution functions. We discuss the possible relationship of this term with the
Sivers function.
\end{abstract}
\date{\today}
\maketitle

\section{Introduction}

The investigation of the quark-gluon dynamics based on both perturbative and
nonperturbative methods is one of the most important subjects of
hadron phenomenology. Recently, due to the new kind of accelerators,
different exclusive hard reactions are begun to be available for study.
The special interests are related to the study of generalized parton distributions
(GPDs), double distributions (DD) and (generalized) distribution amplitudes (GDAs, DAs)  which
are most useful to extract new informations on the composite hadron structure
 \cite{DM, Ji:1996nm, Radyushkin:1997ki, Ball:1996tb, Braun:1999te, Diehl:2003ny, Belitsky:2005qn}.

In \cite{Teryaev:2001qm},
the relations between the GPDs and DDs in the context
of both the direct and inverse Radon transforms have been studied for the first time.
This finding becomes extremely useful for the further investigations of
the Radon transform applications, see for example \cite{Muller:2017wms},
and \cite{HMud1, HMud2, Ch-Th, HM-Th} for the recent studies on the different aspects of the inverse Radon transforms
in terms of GPDs.

The Radon transformations have a long history of application in different
fields: tomography, astronomy etc. Special attempts have been directed
to the inverse problem, that is how to reconstruct the function from its projections.
One of the main problems is associated with the theorem
which states that any function with restricted and compact support can be
uniquely determined or reconstructed by the infinite set of projections only
(see, for example, \cite{Deans, GGV}). This is at odds with the practical use where
we deal with only the finite set of projections.
For the mathematical community, the mentioned ill-posedness of the inverse Radon transform is
closely related to this fact. In contrast, in the present paper, we focus on
the physical aspects of the inverse Radon transformation problem in the context of
GPDs (we refer the readers to \cite{Muller:2017wms} for the modern recent
studies).

The content of the present paper is as follows.
We revisit the (inverse) Radon transforms which connect the
DDs with GPDs. We have found a new term which, as we show, is essentially related to
the $T$-odd $k_\perp$-dependent parton distributions. We demonstrate that
the restrictions applied on the support of the DDs may lead to
the corresponding restrictions for the support of GPDs.

The paper is logically divided into two parts:
the first part is devoted to the rather mathematical abstract issues related to the
(inverse) Radon transforms. We also discuss the problem of the new term existence.
The second part is concentrated on the direct physical application of the
new contribution to the process descriptions through the
transverse-momentum dependent parton distributions.

\section{Getting started: the Fourier and Radon transforms}
\label{Sec1}

For pedagogical reasons, we shortly remind some important definitions and
relations used for the Radon transforms.
All details regarding the different aspects of the Radon transforms can be found in \cite{Deans}.
Also, in this section, we consider the general aspects of the (inverse) Radon transforms for
the unbounded, restricted and localized support of the function $f$ which is a base of the Fourier and Radon transformations.

\subsection{The case of unbounded support of homogeneous function $f$}
\label{SubSecA}

Let us begin with the Fourier transform which helps us
to introduce the Radon transform. For an arbitrary two-dimensional homogeneous function $f(\vec{\bf x})$,
the direct Fourier transform takes the form of
\begin{eqnarray}
\label{F-t-1-dir}
\mathcal{F}[f](\vec{\bf q})= \int_{-\infty}^{+\infty} d^2 \vec{\bf x} \, e^{-i\langle\vec{\bf q},\vec{\bf x}\rangle} \,f(\vec{\bf x}).
\end{eqnarray}
Notice that, in general, both $\mathcal{F}[f](\vec{\bf q})$ and $f(\vec{\bf x})$ can be the complex functions,
{\it i.e.} $\{ \mathcal{F}[f](\vec{\bf q}),\, f(\vec{\bf x})\}\in \mathds{C}$.

As the first step, we assume the support of function $f(\vec{\bf x})$ to be unbounded.
It is convenient to choose the polar coordinates for the vector
$\vec{\bf q}$ ($\vec{\bf q}=\lambda \vec{\bf n}_\varphi$ with the normal vector
$\vec{\bf n}_\varphi=(\cos\varphi, \sin\varphi)$, $\varphi\in [0;\,2\pi]$,
and $\lambda=|\vec{\bf q}|\in [0;\,+\infty]$).
Inserting the integral unit
(the normalization constants have been absorbed into the corresponding integration measures),
\begin{eqnarray}
\label{Int-Unit}
\int_{-\infty}^{+\infty}(dt) \,
\delta\left( t - \langle \vec{\bf q}, \vec{\bf x}\rangle\right)=1,
\end{eqnarray}
we can rewrite the eqn.~(\ref{F-t-1-dir}) as
\begin{eqnarray}
\label{F-t-2-dir}
&&\mathcal{F}[f](\lambda\vec{\bf n}_\varphi)=
\int_{-\infty}^{+\infty} d^2 \vec{\bf x} \,
e^{-i\lambda\langle \vec{\bf n}_\varphi, \vec{\bf x}\rangle}\, f(\vec{\bf x})\times
\nonumber\\
&&
\int_{-\infty}^{+\infty}(dt) \,
\delta\left( t - \lambda\langle \vec{\bf n}_\varphi, \vec{\bf x}\rangle\right).
\end{eqnarray}

Making used the replacement: $\tau=t/\lambda$, we get
\begin{eqnarray}
\label{F-t-3-dir}
\mathcal{F}[f](\lambda\vec{\bf n}_\varphi)\equiv \mathcal{F}[f](\lambda, \varphi)=\int_{-\infty}^{+\infty}(d\tau) \,e^{-i\lambda\tau}\,
\mathcal{R}[f](\tau,\varphi),
\end{eqnarray}
where
\begin{eqnarray}
\label{F-t-4-dir}
\mathcal{R}[f](\tau,\varphi)=
\int_{-\infty}^{+\infty} d^2 \vec{\bf x} \, f(\vec{\bf x})
\delta\left( \tau - \langle\vec{\bf n}_\varphi, \vec{\bf x}\rangle\right)
\end{eqnarray}
defines the direct Radon transformation of $f(\vec{\bf x})$.
If we now perform the rotation of the co-ordinate system as
\begin{eqnarray}
\label{Rot-Sys}
\left\{
  \begin{aligned}
  p&=x_1 \cos\varphi + x_2 \sin\varphi\\
  s&=- x_1 \sin\varphi + x_2 \cos\varphi,\\
  \end{aligned}
\right.
\end{eqnarray}
the eqn.~(\ref{F-t-4-dir}) takes the following form
\begin{eqnarray}
\label{F-t-4-dir-w}
\mathcal{R}[f](\tau,\varphi)&=&
\int dp \, ds \, \delta(\tau - p )
\nonumber\\
&&\times f(p\cos\varphi - s\sin\varphi, p\sin\varphi + s\cos\varphi)
\nonumber\\
&=&\int_{L(\tau, \varphi)} ds \, f(x_1(\tau, \varphi; s), x_2(\tau, \varphi; s)),
\end{eqnarray}
where $L(\tau, \varphi)$ denotes the line given by $\tau - \langle\vec{\bf n}_\varphi, \vec{\bf x}\rangle =0$;
the new co-ordinate $s$ is pointed along the line while the new co-ordinate $p$ is perpendicular to the line.
Hence, one can see that the Radon transform can also be defined through the line integration as
\begin{eqnarray}
\label{R-line-int}
\mathcal{R}[f](\tau,\varphi)=
\int_{L(\tau, \varphi)} d s(\vec{\bf x}) \, f(\vec{\bf x}).
\end{eqnarray}
The eqn.~(\ref{F-t-3-dir}) refers to {\it the Fourier slice theorem} \cite{Deans} and
it means that the one-dimensional Fourier image of the Radon transform of $f$ with respect to the radial (offset)
parameter gives the two-dimensional Fourier transform of $f$, while the angular (slope) parameter $\varphi$ remains untouched.
In other words, the angular parameter $\varphi$
plays the role of both the polar coordinate in $q$-plane and the Radon slope parameter.
%

For our further purposes, it also useful to
introduce an alternative notation for the Radon transform as
\begin{eqnarray}
\label{R-t-1}
\mathcal{R}[f](\tau,\varphi)\equiv {\cal R}_{\tau,\,\varphi}  \left[f(\vec{\bf x})\right],
\end{eqnarray}
which has been used in Sec.~\ref{Sec3}.

From the eqn.~(\ref{F-t-4-dir}), we can see that the Radon transform possesses the symmetry
property in the form of
\begin{eqnarray}
\label{Rad-Sym-1}
\mathcal{R}[f]\left((-)^k \tau,\varphi\right)= \mathcal{R}[f]\left(\tau,\varphi+k\pi\right), \quad
k\in\mathds{Z}.
\end{eqnarray}
Taking into account the symmetry property (\ref{Rad-Sym-1}), one can also write down that
\begin{eqnarray}
\label{F-t-3-dir-2}
&&\mathcal{F}[f](\lambda, \varphi)=
\\
&&\int_{0}^{+\infty}(d\tau) \,\Big\{ e^{-i\lambda\tau}\,
\mathcal{R}[f](\tau,\varphi)
+ e^{i\lambda\tau}\,
\mathcal{R}[f](\tau,\varphi+\pi) \Big\}.
\nonumber
\end{eqnarray}
We stress that the Radon transform $\mathcal{R}[f](\tau,\varphi+\pi)$ can be different from
the Radon transform $\mathcal{R}[f](\tau,\varphi)$ depending on the basic properties of function $f$.

As the angular parameter $\varphi$ is the same for both the Fourier and Radon transforms,
a single full rotation in $q$-plane, $0\leq\varphi\leq 2\pi$, to cover some (unbounded or bounded) domain generates the two rotations in $x$-plane
for the Radon transform.

To derive the inverse Radon transform, at first we invert the Fourier transform (\ref{F-t-1-dir}). We have
\begin{eqnarray}
\label{Inv-F-t-1}
&&f(\vec{\bf x}) = \int_{-\infty}^{+\infty} d^2 \vec{\bf q}\,  e^{+i\langle\vec{\bf q},\vec{\bf x}\rangle} \,\mathcal{F}[f](\vec{\bf q})=
\nonumber\\
&&
\int_{0}^{+\infty} d\lambda \lambda \int_{0}^{2\pi} d\varphi\, e^{+i\lambda\langle \vec{\bf n}_\varphi, \vec{\bf x}\rangle}\,
\mathcal{F}[f](\lambda, \varphi)
\end{eqnarray}
where the polar coordinates have been used again. Using the relation (\ref{F-t-3-dir}) and performing the replacement:
$\tau-\langle \vec{\bf n}_\varphi, \vec{\bf x}\rangle=\eta$, the eqn.~(\ref{Inv-F-t-1}) can be presented as
\begin{eqnarray}
\label{Inv-F-t-2}
&&f(\vec{\bf x}) =
\int_{0}^{+\infty} d\lambda \lambda \int_{0}^{2\pi} d\varphi\times
\nonumber\\
&&
\int_{-\infty}^{+\infty} (d\eta)\, e^{-i\lambda\eta}\,
\mathcal{R}[f](\eta + \langle \vec{\bf n}_\varphi, \vec{\bf x}\rangle, \varphi).
\end{eqnarray}
Given that it remains to implement the integration over the variable $\lambda$ in the eqn.~(\ref{Inv-F-t-2}).
To do this, we observe that the integration variable $\lambda$ can be traded for the derivative over $\eta$ which acts on the exponential function:
\begin{eqnarray}
\label{repl}
\lambda e^{-i\lambda\eta} = i\frac{\partial}{\partial\eta} e^{-i\lambda\eta}.
\end{eqnarray}
Moreover, the integration over $\lambda$, which is singled out, should be regularized by $\eta\to\eta - i\epsilon$:
\begin{eqnarray}
\label{int-lam}
&&i\frac{\partial}{\partial\eta} \int_{0}^{+\infty} d\lambda \, e^{-i\lambda(\eta-i\epsilon)}\equiv 2\pi\frac{\partial}{\partial\eta} \delta_-(\eta)
\nonumber\\
&&=
-\frac{\mathcal{P}}{\eta^2} + i\pi \frac{\partial}{\partial\eta}\delta(\eta).
\end{eqnarray}
Thus, we finally derive for the inverse Fourier transform expressed through the Radon transform:
\begin{eqnarray}
\label{Inv-F-t-3}
&&f(\vec{\bf x}) = f_A(\vec{\bf x}) + f_S(\vec{\bf x})=
\nonumber\\
&&
-i\pi \int_{-\infty}^{+\infty} (d\eta)\int_{0}^{2\pi} d\varphi
\Big[ \frac{\partial}{\partial\eta}\mathcal{R}[f](\eta + \langle \vec{\bf n}_\varphi, \vec{\bf x}\rangle, \varphi) \Big] \delta(\eta)
\nonumber\\
&&-
\int_{-\infty}^{+\infty} (d\eta)\, \frac{\mathcal{P}}{\eta^2}\,
\int_{0}^{2\pi} d\varphi \, \mathcal{R}[f](\eta + \langle \vec{\bf n}_\varphi, \vec{\bf x}\rangle, \varphi).
\end{eqnarray}
Notice that the second term in the eqn.~(\ref{Inv-F-t-3}) is a standard one,
while the first term in the eqn.~(\ref{Inv-F-t-3}) looks, in the most general case, artificial due to the symmetry property of the Radon transformation
provided that {\it (a)} the $\eta$-integration region is (symmetric-)unbounded and {\it (b)}
the angular variable $\varphi$ has been varied in the full region between $0$ and $2\pi$.

Indeed, using the eqn.~(\ref{F-t-4-dir}), we can rewrite the eqn.~(\ref{Inv-F-t-3}) in the equivalent form as
(if $f$ is a regular function)
\begin{eqnarray}
\label{Inv-F-t-3w}
&&f(\vec{\bf x}) =f_A(\vec{\bf x}) + f_S(\vec{\bf x})=
\nonumber\\
&&
-i\pi \int_{0}^{2\pi} d\varphi \int_{-\infty}^{+\infty} (d^2\vec{\bf y}) \,
\big[ \langle\vec{\bf n}_\varphi, \vec{\nabla}\rangle  f(\vec{\bf y})\big]
\delta\left( \langle\vec{\bf n}_\varphi, (\vec{\bf x} - \vec{\bf y})\rangle \right)
\nonumber\\
&&-\int_{0}^{2\pi} d\varphi \int_{-\infty}^{+\infty} (d^2\vec{\bf y}) \,
f(\vec{\bf y})\, \frac{{\cal P}}{\big[ \langle\vec{\bf n}_\varphi, (\vec{\bf x} - \vec{\bf y})\rangle \big]^2}.
\end{eqnarray}
Hence, it becomes clear that the first term in the eqn.~(\ref{Inv-F-t-3}) merely disappears
when we split the angular integration into two parts:
\begin{eqnarray}
\label{phi-1}
\int_{0}^{2\pi} d\varphi = \int_{0}^{\pi} d\varphi + \int_{\pi}^{2\pi} d\varphi.
\end{eqnarray}.

It is instructive to present an alternative way how to derive the inverse Radon transform.
With a help of (\ref{phi-1}) and after the replacements given by $\varphi=\phi + \pi$ and $\lambda= - \tilde\lambda$,
the eqn.~(\ref{Inv-F-t-1}) takes the following form
\begin{eqnarray}
\label{Inv-F-t-1-II}
f(\vec{\bf x}) =
\int_{0}^{\pi} d\varphi \int_{-\infty}^{+\infty} d\lambda \lambda \varepsilon(\lambda) \,
 e^{+i\lambda\langle \vec{\bf n}_\varphi, \vec{\bf x}\rangle}\,
\mathcal{F}[f](\lambda, \varphi).
\end{eqnarray}
We now use the integral representation of $\varepsilon$-function:
\begin{eqnarray}
\label{eps-1}
\varepsilon(\lambda)=\frac{1}{\pi i} \int_{-\infty}^{+\infty} d\rho e^{+i\lambda\rho}\frac{{\cal P}}{\rho}
\end{eqnarray}
and perform two consecutive replacements: the first replacement $\rho + \langle \vec{\bf n}_\varphi, \vec{\bf x}\rangle = \eta$
has to be implemented just after inserting (\ref{eps-1}) into (\ref{Inv-F-t-1-II}) and the second replacement
$\tau - \eta = \tilde\tau$ after using of the representation (\ref{F-t-3-dir}) in eqn.~(\ref{Inv-F-t-1-II}).
Hence, we have
\begin{eqnarray}
\label{Inv-F-t-3-II}
&&f(\vec{\bf x}) =
2 \int_{0}^{\pi} d\varphi \int_{-\infty}^{+\infty} (d\eta)\,
\frac{\mathcal{P}}{\eta - \langle \vec{\bf n}_\varphi, \vec{\bf x}\rangle}\,
\nonumber\\
&&\times
\int_{-\infty}^{+\infty} d\tau
 \, \mathcal{R}[f](\eta + \tau, \varphi) \frac{\partial}{\partial\tau}\delta(\tau)
\end{eqnarray}
and after some simple algebra we get
\begin{eqnarray}
\label{Inv-F-t-3-II-2}
f(\vec{\bf x}) =
- 2 \int_{0}^{\pi} d\varphi \int_{-\infty}^{+\infty} (d\eta)\,
\frac{\mathcal{P}}{\left(\eta - \langle \vec{\bf n}_\varphi, \vec{\bf x}\rangle\right)^2}\,
 \, \mathcal{R}[f](\eta, \varphi).
\end{eqnarray}
Comparing the eqns.~(\ref{Inv-F-t-3}) and (\ref{Inv-F-t-3-II-2}), one can see that
if, in the eqn.~(\ref{Inv-F-t-3}), we split the angular integration as in the eqn.~(\ref{phi-1}), after the consecutive replacements of
integration variables we reproduce the eqn.~(\ref{Inv-F-t-3-II-2}).
It means that the first additional term of eqn.~(\ref{Inv-F-t-3})
is an insubstantial contribution iff we have the full angular integration in the interval $[0, 2\pi]$ and
the symmetric integration measure of the radial $\eta$-parameter.

It is worth to notice that the representation (\ref{Inv-F-t-3-II-2}) is fully coincided with the
result which can be found in \cite{Deans, GGV} where the integration over the full angular region, $\varphi\in [0,\,2\pi]$,
has been used {\it a priori}.

However, as we demonstrate in Sec.~\ref{SubSecB}, there are some cases where for the restricted
support of $f$ the angular parameter $\varphi$
together with the radial parameter $\eta$
meet the corresponding restrictions limiting the variation intervals and
breaking the symmetry condition generated by the eqn.~(\ref{Rad-Sym-1}).
At the same time, as explained in Sec.~\ref{Sec4}, in the case of
the generalized transverse-momentum dependent distributions it turns out that the angular integration is also limited by
the region $\varphi\in\left[-\pi/2, \pi/2\right]$ only owing to the fact that the $T$-reversal invariance has been relaxed.
Hence, we deal with the situation where there is no the symmetry presented by
(\ref{Rad-Sym-1}) as well.
As a result, it is important for the first additional term of eqn.~(\ref{Inv-F-t-3}) to be included in the analysis.

\subsection{The case of restricted support of function $f$}
\label{SubSecB}

In the preceding subsection, we have considered the simplest case of unbounded support of function $f$.
However, in any practical applications of the Radon transforms, as a rule the support of function $f$ is restricted by
some region. In this subsection, we study the influences of the restricted support on the Fourier and
Radon transformations.
Namely, we demonstrate that, in the most general case, the restricted support of $f$ leads to the
restrictions in the $q$-plane where the Fourier image of $f$ has been determined.

To begin with, let us return to the Fourier slice theorem, see the eqn.~(\ref{F-t-3-dir}). Notice that the integral representation
of unit~(\ref{Int-Unit}), which has been inserted into the direct
Fourier transform~(\ref{F-t-1-dir}), can be actually presented in the different forms.
Indeed, we have
\begin{eqnarray}
\label{Int-Unit-2}
&&\int_{-\infty}^{+\infty}(dt) \,
\delta\left( t - \langle \vec{\bf q}, \vec{\bf x}\rangle\right)=
\int_{-\infty}^{+\infty}(d\tau) \,
\delta\left( \tau - \langle \vec{\bf n}_\varphi, \vec{\bf x}\rangle\right)=
\nonumber\\
&&
\int_{-\infty}^{+\infty}(dz) \,
\delta\left( z - x_1 - \xi x_2\right)=
1,
\end{eqnarray}
where
\begin{eqnarray}
\label{z-defs}
&&z= \frac{t}{q_1}\quad \text{with}\quad dz= \frac{dt}{q_1},
\\
\label{xi-defs}
&&\xi= \frac{q_2}{q_1}.
\end{eqnarray}
Then, we write down the direct Fourier transform as
\begin{eqnarray}
\label{Dir-F-t-1}
&&\mathcal{F}[f](q_1,q_2)=
\\
&&
\int_{-\infty}^{+\infty} d^2\vec{\bf x}  \, e^{-i\langle\vec{\bf q},\vec{\bf x}\rangle} \,f(\vec{\bf x})
\int_{-\infty}^{+\infty}(dz) \,
\delta\left( z - x_1 - \xi x_2\right) \Big|_{\xi=\frac{q_2}{q_1}}=
\nonumber\\
&&
\int_{-\infty}^{+\infty}(dz) \,\, e^{-iq_1 z} \Big\{
\int_{-\infty}^{+\infty} d^2\vec{\bf x} \,f(\vec{\bf x})\,
\delta\left( z - x_1 - \xi x_2\right)\Big\} \Big|_{\xi=\frac{q_2}{q_1}}.
\nonumber
\end{eqnarray}
Hence, the Fourier slice theorem takes now the following form
\footnote{
Notice also that it is sometimes useful to present the Radon transform in the form of
$
\mathcal{R}[f](\tau, \varphi) = \mathcal{R}[f](z, \xi)/|\cos\varphi|,
$
where
$
z=\tau/\cos\varphi,\quad \xi=\tan\varphi\equiv q_2/q_1.
$
}
:
\begin{eqnarray}
\label{F-s-th}
\int_{-\infty}^{+\infty}(dz) \,e^{-iq_1 z}\,
\mathcal{R}[f](z,\xi) = \mathcal{F}[f](q_1, \xi q_1)\Big|_{\xi=\frac{q_2}{q_1}}.
\end{eqnarray}
In contrast to the eqn.~(\ref{F-t-3-dir}), this representation deals with $(q_1, \xi)$ as an independent set of variables for the Fourier transform.
As a result, the $\delta$-function in the direct Radon transform, $\delta\left( z - x_1 - \xi x_2\right)$, gives
$(z,\xi, x_2)$ as the independent variables, {\it i.e.}
\begin{eqnarray}
\label{R-t-L}
\mathcal{R}[f](z,\xi)= \int_{-\infty}^{+\infty}dx_2 \, f(z-\xi x_2, x_2).
\end{eqnarray}

So far, we did not impose any support restrictions for the function $f$.
Now, for the sake of simplicity, let $\vec{\bf x}$ be in the region $[-1,\, 1]$, {\it i.e.} $\vec{\bf x}\in\Omega_\Box$,
and let $f$ be an homogeneous functions. Regarding the function support,
this example is very close to the situation which appears in the case of GPDs.

We want to show that the support restrictions applied for the function $f$ lead to the corresponding restrictions
for the support of function $\mathcal{F}[f](\vec{\bf q})$ in $q$-plane and, therefore, to the restrictions for
the (inverse) Radon transformations, see the eqn.~(\ref{F-t-3-dir}).

Owing to $z-x_1-\xi x_2=0$, which is an argument of the $\delta$-function in the Radon transform (\ref{Dir-F-t-1}),
one can readily derive that
\begin{eqnarray}
\label{Res-0-1}
\text{if}\,\,\, x_1\in[-1,\,1], \,\,\,\text{then}\,\,\,
z+1\geq \xi x_2 \geq z-1.
\end{eqnarray}
Since $x_2\in[-1, 1]$, the given inequality generates four different inequalities depending
on the position of $(z\pm 1)/\xi$ in the interval $[-1, 1]$, {\it i.e.}
\begin{eqnarray}
\label{ineq-con}
&&(A): \quad \frac{z+1}{\xi}\,\,\text{and}\,\,\frac{z-1}{\xi}\, \in [-1, 1];
\\
&&(B): \quad \frac{z+1}{\xi}\,\, \in [-1, 1], \,\,\,\frac{z-1}{\xi}\, \slashed{\in} [-1, 1];
\\
&&(C): \quad \frac{z+1}{\xi}\,\, \slashed{\in} [-1, 1],\,\,\,\frac{z-1}{\xi}\, \in [-1, 1];
\\
&&(D): \quad \frac{z+1}{\xi}\,\, \slashed{\in} [-1, 1],\,\,\,\frac{z-1}{\xi}\, \slashed{\in} [-1, 1].
\end{eqnarray}
However, for our demonstration, it is enough to consider the case $(A)$.
So, we dwell on the following inequalities
\begin{eqnarray}
\label{Res-1}
1\stackrel{\text{a}}{\geq}\frac{z+1}{\xi}\geq x_2 \geq \frac{z-1}{\xi} \stackrel{\text{b}}{\geq} -1.
\end{eqnarray}

Since we want to study the support restrictions in the Cartesian system,
the next steps are the following: {\it (a)} we replace the angular (slope) parameter $\xi$ in (\ref{Res-1}) by
the Cartesian variables $(q_1, q_2)$ using the eqn.~(\ref{xi-defs});
{\it (b)} we consider $q_2$ as a function of $q_1$ and $z$ (here, $z$ plays a role of an external parameter
for this kind of function) in $q$-plane.

For our convenience, we also introduce the following regions in $q$-plane:
\begin{eqnarray}
\label{Reg-q}
&\Omega^{I}&: \quad \{q_2>0,  q_1>0\} \doteq \vec{\bf q}^{I};
\nonumber\\
&\Omega^{II}&: \quad \{q_2<0,  q_1>0\}\doteq \vec{\bf q}^{II};
\nonumber\\
&\Omega^{III}&: \quad \{q_2<0,  q_1<0\}\doteq \vec{\bf q}^{III};
\nonumber\\
&\Omega^{IV}&: \quad \{q_2>0,  q_1<0\}\doteq \vec{\bf q}^{IV};
\end{eqnarray}
\begin{figure}[t]
\centerline{\includegraphics[width=0.5\textwidth]{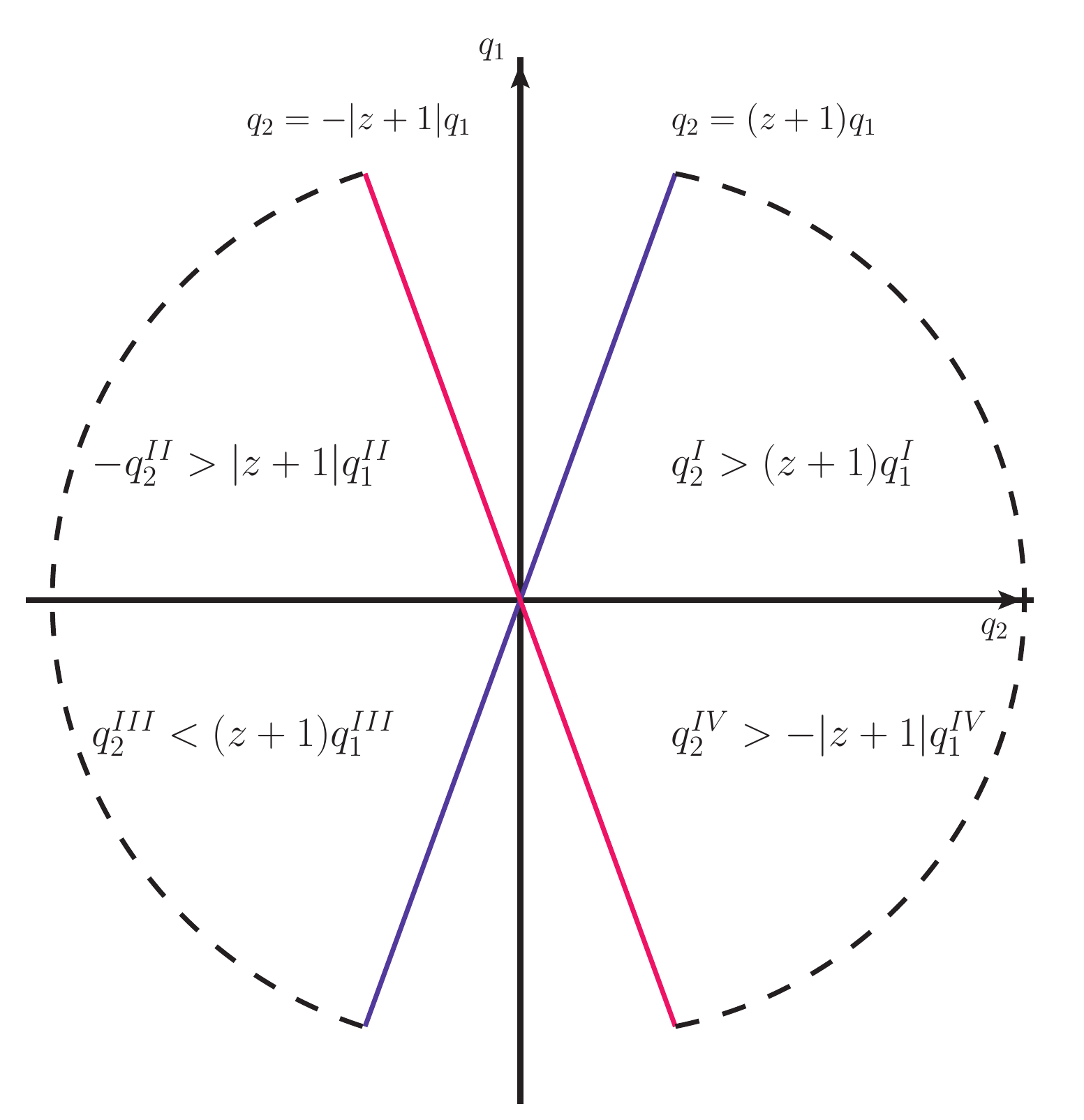}}
\caption{The $\Omega$-regions in $q$-plane for the Fourier function support.}
\label{Fig-q}
\end{figure}

Given that, for the definiteness, we assume $\xi$ and $z+1$ to be the positive values, {\it i.e.} $\{\xi,\,\, z+1\} > 0$.
Hence, the condition $a$ of (\ref{Res-1}) together with (\ref{xi-defs}) give us that
\begin{eqnarray}
\label{Res-q-1}
(z+1) q_1^I \leq q_2^I\,\,\, \text{and}\,\,\, (z+1) q_1^{III} \geq q_2^{III}.
\end{eqnarray}
 At the same time, the case of the negative values for $\xi$ and $z+1$ corresponds to the following conditions:
\begin{eqnarray}
\label{Res-q-2}
|z+1| q_1^{II} \leq - q_2^{II}\,\,\, \text{and}\,\,\, - |z+1| q_1^{IV} \leq q_2^{IV}.
\end{eqnarray}

We emphasize that the equalities in (\ref{Res-q-1}) and
(\ref{Res-q-2}) like $(z+1) q_1^I = q_2^I$ etc. correspond finally to the requirement which reduces a number
of independent variables. Since, from the very beginning, we have considered $f$ as a function of two
independent variables, then the boundaries defined by the equalities in (\ref{Res-q-1}) and (\ref{Res-q-2})
should be excluded from the domain.

In particular cases, these statements are known as the following theorem in the mathematical literature:
{\it if $f$ is a function of $n$ independent variables then the Radon
transform of $f$, $\mathcal{R}[f]$, depends on $n$ independent variables too.
That is, the Radon transform is a bijection and lives on
$\mathbb{R}^1\times \mathrm{S}^{n-1}$} \cite{Deans}.

To conclude, the inequalities (without the equalities) of (\ref{Res-q-1}) and (\ref{Res-q-2})
define the so-called physical domain depicted in Fig.~\ref{Fig-q} where
the two-dimensional Fourier and, therefore, the Radon transforms have been uniquely determined for the
function $f$ with the restricted support: $\text{supp}(f)=\Omega_\Box$.
This ultimately means that the function $f$ can be reconstructed by the inverse Radon transform iff
the support of Fourier function $\mathcal{F}[f](\vec{\bf q})$ in $q$-plane has been compactly distributed
along $q_2$-axis within the right or/and left $\Omega$-regions, see Fig.~\ref{Fig-q}.

In other words, the
correct inverse Radon transform should be implemented with the angular parameter $\varphi$ which is
limited by the region $[0, \pi]$ (or, after shifting, by $[-\pi/2, \pi/2]$) corresponding to Fig.~\ref{Fig-q}.
{\it This allows the first term in the eqn.~(\ref{Inv-F-t-3})
to exist and to do contribute to the inversion of Radon transform in the case of
$ f(\vec{\bf x})\in \mathds{C}$}.

\subsection{The case of localized support of function $f$}
\label{SubSecC}

The same conclusions on the restrictions in the Cartesian system can be demonstrated by
the direct calculation of the
Radon image of the simple localized function. Indeed, if we suppose that the localized function
has been defined as (here, $\vec{\bf A}$ is an arbitrary external vector)
\begin{eqnarray}
\label{Loc-f-1}
f(\vec{\bf x} \,| \vec{\bf A}) = e^{-(\vec{\bf x} - \vec{\bf A})^2},
\end{eqnarray}
the direct Radon image of this function reads
\begin{eqnarray}
\label{R-Loc-f-1}
\mathcal{R}[f](\tau,\varphi\, | \vec{\bf A})= \int_{-\infty}^{+\infty} d^2\vec{\bf x}\, \,e^{-(\vec{\bf x} - \vec{\bf A})^2} \,
\delta\left( \tau - \langle\vec{\bf n}_\varphi, \vec{\bf x}\rangle\right).
\end{eqnarray}
After direct calculations, we get (here, $C$ is a constant)
\begin{eqnarray}
\label{R-Loc-f-2}
\mathcal{R}[f](\tau - \langle\vec{\bf n}_\varphi, \vec{\bf A}\rangle ,\varphi)= C \,\,
e^{-(\tau - \langle\vec{\bf n}_\varphi, \vec{\bf A}\rangle)^2}.
\end{eqnarray}
Indeed, after the replacement $\vec{\bf y}=\vec{\bf x} - \vec{\bf A}$ in eqn.~(\ref{R-Loc-f-1}),
which leads to 
\begin{eqnarray}
\label{expl-1}
&&\mathcal{R}[f](\tau,\varphi \, | \vec{\bf A})= \int_{-\infty}^{+\infty} d^2\vec{\bf y}\, \,e^{-\vec{\bf y}^2} \,
\delta\left( \tau - \langle\vec{\bf n}_\varphi, \vec{\bf A}\rangle - \langle\vec{\bf n}_\varphi, \vec{\bf y}\rangle\right)
\nonumber\\
&& = \mathcal{R}[f](\tau - \langle\vec{\bf n}_\varphi, \vec{\bf A}\rangle ,\varphi)\equiv
\mathcal{R}[f](\widetilde\tau,\varphi),
\end{eqnarray}
we perform the rotation of the system, see eqn.~ (\ref{Rot-Sys}), to obtain
the following representation
\begin{eqnarray}
\label{expl-1}
&&
\mathcal{R}[f](\widetilde\tau,\varphi)=
\int_{-\infty}^{+\infty} dp \,ds \, \,e^{-p^2-s^2} \,
\delta\left( \widetilde\tau - p\right)
\nonumber\\
&&=e^{-\widetilde\tau^2} \,\int_{-\infty}^{+\infty} ds \, \,e^{-s^2}
=C \,
e^{-(\tau - \langle\vec{\bf n}_\varphi, \vec{\bf A}\rangle)^2}.
\end{eqnarray}
Then, we replace the polar Radon co-ordinates $(\tau, \varphi)$ by the Cartesian co-ordinates
$(u, v)$ as
\begin{eqnarray}
\label{C-R-sys}
&&\tau=\sqrt{\vec{\bf K}^2}, \quad \vec{\bf K}= (u,\, v)
\nonumber\\
&&
u= \tau\, \cos\varphi, \quad v=\tau\, \sin\varphi.
\end{eqnarray}
With these, we derive that
\begin{eqnarray}
\label{R-Loc-f-2}
\mathcal{R}[f](u, v | \vec{\bf A})= C \,\,
e^{- \frac{\vec{\bf K}^2 - \langle\vec{\bf K}, \vec{\bf A}\rangle}{\sqrt{\vec{\bf K}^2}}},
\end{eqnarray}
which is localized in the Cartesian system of co-ordinates $(u, v)$.

\section{The Radon transforms for GPDs and DD functions: the support theorem and
the extra GPDs properties}
\label{Sec4}

As stated above, we have considered the (inverse) Radon transforms defined for an
arbitrary functions $f$ with the unbounded and compact supports.
We have demonstrated that the restricted support of $f$ defined as $\text{supp}(f)=\Omega_\Box$
leads to the restriction imposed on the angular Radon parameter $\varphi\in [-\pi/2, \, \pi/2]$.
Now, based on the results presented in the preceding sections,
we overhaul the (inverse) Radon transforms which relate
the GPDs to DDs (cf. \cite{Teryaev:2001qm}).
Since, as well-known, the support of DD-function is similar to the considered region $\Omega_\Box$
(modulo the corresponding rotation and re-scaling), we should expect that the similar restriction on
the angular Radon parameter takes place also in the case of GPDs independently
of the presence of the primordial transverse momenta.
We remind that the $k_\perp$-dependence of the parton distributions results in the
relaxing of $T$-reversal invariance and, therefore, the DD-functions for the positive
and negative variable $x_2$ (or, in the standard notation, $\beta$, see Fig.~\ref{Fig-S})
start to be independent ones, see Sec.~\ref{Sec3}.

In the first part of this section, we show that the restricted support of the DD-function in the
physical domain can conditionally lead to the physical restrictions for the GPD-variables $z$ and $\xi$.
In contrast to the abstract mathematical case considered in the preceding section,
we deal now with the physical interpretation of the angular Radon parameter $\xi$.
Namely, we have both the physical, $\xi \leq 1$ (this is the GPDs region), and unphysical, $\xi > 1$
(this is the GDA region), values of the angular Radon parameter.

In the second part of the section, we study the extra properties of GPDs which come from the
the requirement of $h_A=0$.

\subsection{The support theorem}
\label{B-Q-theorem}

We begin with the standard relation between
GPDs and DDs in the following form \cite{Radyushkin:1997ki}:
\begin{eqnarray}
\label{gpd-dd-1}
H(z, \xi)=
\int_{-\infty}^{+\infty} d\alpha d\beta \, \widetilde h (\alpha, \beta)
\delta\left( z - \alpha - \xi\beta\right),
\end{eqnarray}
where
\begin{eqnarray}
\label{Tilde-h}
&&\widetilde h(\alpha, \beta) = h(\alpha, \beta) \Theta\big(\{\alpha,\beta\}\in\overline{\Omega}\big),
\nonumber\\
&&\int_{\overline{\Omega}} d\alpha d\beta \doteq \int_{-1}^{+1} d\alpha \int_{-1+|\alpha|}^{1-|\alpha|} d\beta .
\end{eqnarray}
As usual, the function $h(\alpha, \beta)$ involves the
standard DD-functions $f(\alpha, \beta)$ and $g(\alpha, \beta)$ \cite{Belitsky:2005qn} but it is irrelevant now.

The full support region $\overline{\Omega}$, which is a symmetric rhombus,
can be split-up into four subregions: $\overline{\Omega}^I, ..., \overline{\Omega}^{IV}$ where,
for example, the first region  $\overline{\Omega}^I$ is given by $0\leq \alpha + \beta \leq 1$ together
with $0\leq \alpha \leq 1$ and $0\leq \beta \leq 1$ and in the similar manner the other regions can be determined.

Let us write the eq.~(\ref{F-t-4-dir}) in the equivalent form as
\begin{eqnarray}
\label{F-t-4-dir-2}
&&\mathcal{R}[f](\eta,\varphi) |\cos\varphi|=
\nonumber\\
&&\int_{\overline{\Omega}} dx_1 dx_2 \, f(x_1, x_2)
\delta\Big( \frac{\eta}{\cos\varphi} - x_1 - x_2\tan\varphi \Big)
\end{eqnarray}
and, as above, we introduce the following relations
\begin{eqnarray}
\label{Dict-1}
&&\mathcal{R}[f](\eta,\varphi) |\cos\varphi| \Rightarrow H\Big(\frac{\eta}{\cos\varphi},\tan\varphi \Big) = H(z, \xi)
\nonumber\\
&&\frac{\eta}{\cos\varphi} \Rightarrow z, \quad \tan\varphi \Rightarrow \xi, \quad f(x_1, x_2) \Rightarrow h(\alpha, \beta).
\end{eqnarray}

As shown in Sec.\ref{Sec1}, the restricted support of any arbitrary function (for example, $\widetilde h(\alpha, \beta)$)
induces the maximal angular restriction $\varphi\in [-\pi/2, \pi/2]$ (or $\xi\in [-\infty, \infty]$)
where the symmetry property presented by the eqn.~(\ref{Rad-Sym-1}) has been broken
leaving the first term in the eqn.~(\ref{Inv-F-t-3}) does contribute, see Sec.~\ref{Sec3}.

Let us be limited by consideration of the region $\overline{\Omega}^I$ which is enough in the most of cases.
If the support of $h(\alpha, \beta)$ is fixed and defined by $\overline{\Omega}^I$,
then a question which is arisen is that whether the support of $H$-function in eqn.~(\ref{gpd-dd-1}) is
automatically giving the region $0 \leq \{z,\, \xi\} \leq 1$.
To cover the region $\overline{\Omega}^I$, the Radon parameters $\eta$ and $\varphi$ have to vary within the
intervals $[0,\,1]$ and $[0,\,\pi/2]$, respectively.
Hence, at the moment we deal with the unbounded region in terms of $z$ and $\xi$, {\it i.e.}
$\{z,\,\xi\}\in [0,\,+\infty]$.

On the other hand, making used the integral representation of the $\Theta$-function defined as
\begin{eqnarray}
\label{Theta-int-1}
&&\Theta\big(\{\alpha,\beta\}\in\overline{\Omega}^I\big)\equiv
\nonumber\\
&&\theta\left( 0\leq \alpha + \beta \leq 1\right)
\theta\left( 0\leq \alpha \leq 1\right)\theta\left( 0\leq \beta \leq 1\right) =
\nonumber\\
&&\int_{0}^{1} dT \delta\left( T- \alpha - \beta\right)\Big|_{0\leq \{\alpha,\beta\}\leq 1},
\end{eqnarray}
we obtain that the {\it r.h.s.} of (\ref{gpd-dd-1}) reads
\begin{eqnarray}
\label{Supp-1}
\int_{0}^{1} dT \, \int_{-\infty}^{+\infty} d\alpha d\beta \, h (\alpha, \beta)
\delta\left( T- \alpha - \beta\right)\, \delta\left( z - \alpha - \xi\beta\right).
\end{eqnarray}
After integrating over $d\alpha$ and $d\beta$ with two $\delta$-functions, we derive that
\begin{eqnarray}
\label{Supp-2}
H(z, \xi)=\frac{1}{|1-\xi|}\int_{0}^{1} dT \, h \Big(\frac{z-\xi T}{1-\xi}, \frac{T-z}{1-\xi}\Big)
\end{eqnarray}
provided
\begin{eqnarray}
\label{Con-1}
0\stackrel{\text{a}}{\leq} \frac{z-\xi T}{1-\xi} \stackrel{\text{b}}{\leq} 1,
\quad 0\stackrel{\text{c}}{\leq} \frac{T-z}{1-\xi}\stackrel{\text{d}}{\leq} 1.
\end{eqnarray}
Consider the condition $a$ of (\ref{Con-1}), we have
\begin{eqnarray}
\label{con-a}
&&\text{if}\,\,\, 1-\xi\geq 0, \,\, \text{then}\,\, z\geq\xi T,
\nonumber\\
&&\text{if}\,\,\, 1-\xi\leq 0, \,\, \text{then}\,\, z\leq\xi T.
\end{eqnarray}
Hence, it means that $z,\xi$ can be unbounded.
At the same time, focusing on the condition $b$ of (\ref{Con-1}), we have
\begin{eqnarray}
\label{con-b}
&&\text{if}\,\,\, 1-\xi\geq 0, \,\, \text{then}\,\,\xi (T-1)\geq z-1
\nonumber\\
&&\Rightarrow z\leq 1\,\,\,\text{or}\,\,\, z\leq 1+ \xi(T-1),
\nonumber\\
&&\text{if}\,\,\, 1-\xi\leq 0, \,\, \text{then}\,\,\xi (T-1)\leq z-1
\nonumber\\
&&\Rightarrow z\geq 1\,\,\,\text{or}\,\,\, z\geq 1+ \xi(T-1).
\end{eqnarray}
In the same manner, we are able to analyse the conditions $c$ and $d$ of (\ref{Con-1}).
From the condition $c$ of (\ref{Con-1}), we obtain that
\begin{eqnarray}
\label{con-c}
&&\text{if}\,\,\, 1-\xi\geq 0, \,\, \text{then}\,\, T\geq z \Rightarrow z\leq 1,
\nonumber\\
&&\text{if}\,\,\, 1-\xi\leq 0, \,\, \text{then}\,\, T\leq z
\end{eqnarray}
and, the condition $d$ of (\ref{Con-1}) gives us the following
\begin{eqnarray}
\label{con-d}
&&\text{if}\,\,\, 1-\xi\geq 0, \,\, \text{then}\,\, T+\xi\leq 1+ z,
\nonumber\\
&&\text{if}\,\,\, 1-\xi\leq 0, \,\, \text{then}\,\, T+\xi\geq 1+ z.
\end{eqnarray}

\begin{figure}[t]
\centerline{\includegraphics[width=0.4\textwidth]{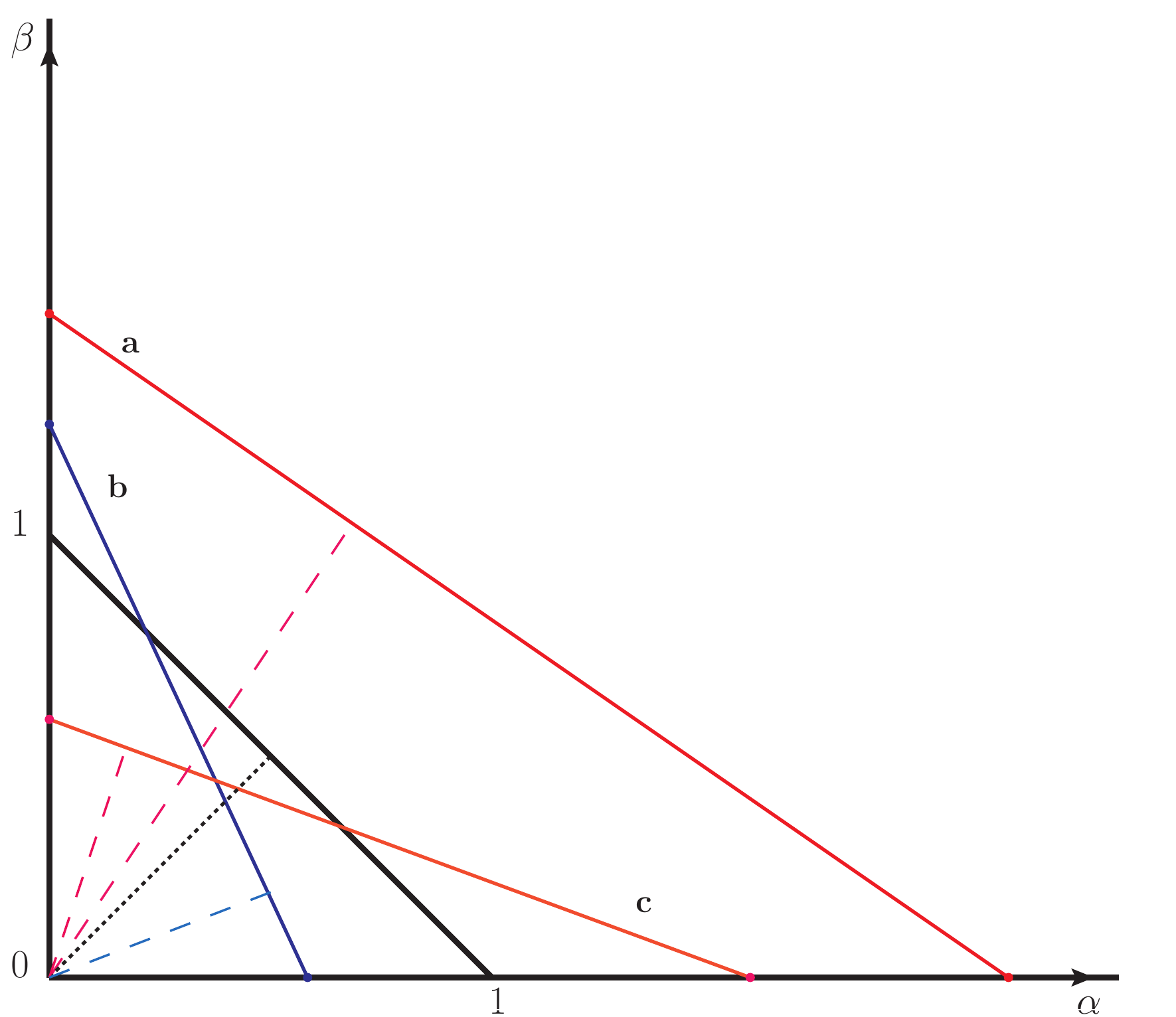}}
\caption{The $\overline{\Omega}^I$ region of the DD-function support and
the three classes of lines, $\beta=z/\xi - \alpha/\xi$,
related to the corresponding Radon transforms. Notations:
${\bf a}:\Rightarrow \{z/\xi>1 (\,\text {or}\,\, z > \xi)\} \,\bigcup \, \{z > 1, \xi > 1 (\,\text {or}\,\, \varphi > \pi/4)\}$;
${\bf b}:\Rightarrow \{z/\xi>1 (\,\text {or}\,\, z > \xi)\} \,\bigcup \, \{z < 1, \xi < 1 (\,\text {or}\,\, \varphi < \pi/4)\}$;
${\bf c}:\Rightarrow \{z/\xi<1 (\,\text {or}\,\, z < \xi)\} \,\bigcup \, \{z > 1, \xi > 1 (\,\text {or}\,\, \varphi > \pi/4)\}$.
}
\label{Fig-S}
\end{figure}

Thus, we can infer that {\it the conditions (\ref{Con-1}) cannot ensure, generally speaking,
the physical restrictions for the parameters $z$ and $\xi$
unless we impose $\xi \leq 1$ from the very beginning}.
Fig.~\ref{Fig-S} demonstrates graphically the above statement and
shows the three most useful classes of the lines.
At the same time, this figure
reflects the well-known situation which can be presented as
\begin{eqnarray}
\label{GPD-GDA}
&&H^{\text{GPD}}(z, \xi)\theta(\xi \leq 1) \oplus H^{\text{GDA}}(z, \xi)\theta(\xi > 1) =
\nonumber\\
&&
\int_{-\infty}^{+\infty} d\alpha d\beta \, \widetilde h (\alpha, \beta)
\delta\left( z - \alpha - \xi\beta\right).
\end{eqnarray}
We can see that the line denoted as ${\bf a}$ corresponds to the unphysical regions for
$z$ and $\xi$ and, according to the Boman-Quinto theorem \cite{BQ-th}, this class of lines can be excluded from the consideration.
The line ${\bf b}$ in Fig.~\ref{Fig-S} corresponds to the physical region for $z$ and $\xi$
(describing, in this case, the DGLAP region for
the GPDs), while the line ${\bf c}$ corresponds to the unphysical region of parameters $z$ and $\xi$ and describes
the GDA.

Hence, to reconstruct the DD-function by the inverse Radon transforms we are forced to deal with the $H$-function which
is supported by the unphysical region $\{\xi > 1, z\in \forall \}$ \cite{Teryaev:2001qm}.
In other words, for $t$-channel, the physical region of $h$-function should also need our knowledge for to the physical
region of $H$-function in $s$-channel. This is one of exhibitions of the ill-posed problem related
to the inverse Radon transforms \cite{Muller:2017wms, Gabdrakhmanov:2019bqm}.
The other formulation of ill-posed problem has been presented in App.~\ref{App1}.

\subsection{The additional term $f_A$}
\label{SubSec:Add-term}

We are now in position to study the properties of the additional term $f_A$ (or $h_A$).
As follows from Sec.~\ref{Sec1}, if we have the angular parameter $\varphi$ which varies in the
full interval between $0$ and $2\pi$ then we can conclude that
\begin{eqnarray}
\label{fA-1}
&&\frac{i}{\pi}f_A(\vec{\bf x}) =
\int_{-\infty}^{+\infty} (d\eta) \, \delta(\eta)\, \int_{0}^{2\pi} d\varphi
\nonumber\\
&&
\times\Big[ \frac{\partial}{\partial\eta}\mathcal{R}[f](\eta + \langle \vec{\bf n}_\varphi, \vec{\bf x}\rangle, \varphi) \Big] =0.
\end{eqnarray}
We can also treat this equation as the condition which GPDs have to fulfil to.
In other words, in order to satisfy the eqn.~(\ref{fA-1}) the corresponding GPDs should have
definite symmetric properties.
Indeed, since the skewness parameter $\xi$ is determined by $\tan\varphi$,
let us first to make replacement given by $\tilde\varphi=\varphi -\pi/2$ in the integrand of eqn.~(\ref{fA-1}), we obtain
\begin{eqnarray}
\label{fA-1-2}
&&\frac{i}{\pi}f_A(\vec{\bf x}) =
\int_{-\infty}^{+\infty} (d\eta) \, \delta(\eta)\,
\Big[ \int_{-\frac{\pi}{2}}^{\frac{\pi}{2}} d\tilde\varphi +  \int_{\frac{\pi}{2}}^{\frac{3\pi}{2}} d\tilde\varphi\Big]
\nonumber\\
&&
\times\Big[ \frac{\partial}{\partial\eta}\mathcal{R}[f](\eta +
\langle \vec{\bf n}_{\tilde\varphi+\frac{\pi}{2}}, \vec{\bf x}\rangle, \tilde\varphi+\frac{\pi}{2}) \Big] =0.
\end{eqnarray}
Here, the Radon transform in the integrand can be presented as
\begin{eqnarray}
\label{Rad-Tr-int}
&&\hspace{-0.4cm}\mathcal{R}[f](\eta +
\langle \vec{\bf n}_{\tilde\varphi+\frac{\pi}{2}}, \vec{\bf x}\rangle, \tilde\varphi+\frac{\pi}{2}) =
\\
&&
\hspace{-0.4cm}\int_{-\infty}^{+\infty} d^2 \vec{\bf y}\, f(y_1, y_2) \,
\delta(\eta -(x_1-y_1) \sin\tilde\varphi + (x_2-y_2) \cos\tilde\varphi)=
\nonumber\\
&&
\hspace{-0.4cm}\int_{-\infty}^{+\infty} d^2 \vec{\bf y}\, f(y_2, y_1) \,
\delta(\eta -(x_1-y_2) \sin\tilde\varphi + (x_2-y_1) \cos\tilde\varphi)=
\nonumber\\
&&
\hspace{-0.4cm}\int_{-\infty}^{+\infty} d^2 \vec{\bf y}\, \frac{f(y_2, y_1)}{|\cos\tilde\varphi|} \,
\delta(\eta + x_2 - \xi x_1 - y_1 + \xi y_2 ).
\nonumber
\end{eqnarray}
Given that, using the eqns.~(\ref{F-t-4-dir-2}) and (\ref{Dict-1}) we readily derive that
\begin{eqnarray}
\label{fA-2}
\hspace{-0.2cm}\frac{\partial}{\partial x_2} \int_{-\infty}^{\infty} d\xi \Big\{
\bar H(x_2-\xi x_1, -\xi) + \bar H(x_2+\xi x_1, +\xi)
\Big\}=0
\end{eqnarray}
where a new function $\bar H$ is introduced to be equal to
\begin{eqnarray}
\label{barH}
\bar H (z, \xi) =
\int_{-\infty}^{+\infty} d^2 \vec{\bf y}\, f(y_2, y_1) \, \delta(z - y_1 - \xi y_2).
\end{eqnarray}
To obtain the eqn.~(\ref{fA-2}),  the derivative over $d\eta$ in the eqn.~(\ref{fA-1-2}) has been replaced by the derivative over $dx_2$ and
the integration over $d\eta$ has been performed.
We stress that in contrast to the standard function $H$, see the eqn.~(\ref{gpd-dd-1}),
the function $\bar H$ introduced in the eqn.~(\ref{barH}) corresponds to the Radon transform of $f(y_2, y_1)$ rather than
$f(y_1, y_2)$.

Therefore, to fulfil the condition (\ref{fA-2}), the GPDs have to obey either the
symmetry given by
\begin{eqnarray}
\label{barH-sym}
\bar H(x_2-\xi x_1, -\xi)= - \bar H(x_2+\xi x_1, \xi)
\end{eqnarray}
or
\begin{eqnarray}
\label{barH-const}
\frac{\partial}{\partial x_2} \bar H(x_2\pm\xi x_1, \pm\xi) = 0.
\end{eqnarray}
The condition (\ref{barH-const}) cannot to be realized because the function $f_A$ is a function of two variables by construction.
With respect to the symmetry defined by (\ref{barH-sym}), it can be considered as the symmetry properties
for the standard GPDs $H$
under the time-reverse transforms provided that the function $f$ satisfies the requirement
\begin{eqnarray}
\label{f-sym}
f(y_1, y_2) = \pm f(y_2, y_1).
\end{eqnarray}

Notice that if we deal with the case where the time-reverse invariance is relaxed that we have no such GPDs which
obey the condition (\ref{barH-sym}) and, therefore, the additional term $f_A$ exists.

\section{Additional term $h_A$ and Generalized transverse-momentum dependent functions}
\label{Sec3}

In this section, we give a possible physical interpretation of the additional term $f_A$ discussed in the
previous sections. Since one of the important features of $f_A$ is its complexness, our interpretation is based on the
generalized transverse-momentum dependent functions (GTMD-functions) where
the imaginary parts of GTMD exist due to the fact that the time-reversal invariance has been relaxed
(below, we give the exact definition of the time-reversal relaxing).
To this end, we also introduce the DD-functions which involve the transverse momentum $k_\perp$ dependence.

We begin with the momentum dependent functions which can be introduced as
(within the Heisenberg representation)
\begin{eqnarray}
\label{gtmdf-1}
W^{[\Gamma]}_{s_1\,s_2}(k; \overline{P}, \Delta) =
\int(d^4 z) e^{+iK\cdot z} \langle p_2, s_2|
{\cal O}^{[\Gamma]}(0; z) | p_1, s_1 \rangle,
\end{eqnarray}
where
\begin{eqnarray}
\label{Op}
&&\hspace{-0.5cm}{\cal O}^{[\Gamma]}(0; z)=: \bar\psi(0) \Gamma \, \left[0;\,z \right]_A \psi(z):
\\
&&\hspace{-0.5cm} K=k-\frac{\Delta}{2}, \, \overline{P}=\frac{p_2 + p_1}{2}, \, \Delta=p_2 - p_1, \, -2\xi=\frac{\Delta^+}{\overline{P}^+}.
\end{eqnarray}
We now go over to the so-called fractional projections which appears as a result of factorization, we have
\begin{eqnarray}
\label{gtmdf-2}
&&W^{[\Gamma]}_{s_1\,s_2}(x, \vec{\bf k}_\perp; \overline{P}, \Delta)
=
\nonumber\\
&&\int(dk^-)(dk^+) \delta(x - k^+/\overline{P}^+) W^{[\Gamma]}_{s_1\,s_2}(k; \overline{P}, \Delta)=
\\
&&\int(dz^-)(d^2\vec{\bf z}_\perp) e^{+iK^+ z^- -i\vec{\bf k}_\perp\vec{\bf z}_\perp }
\langle p_2, s_2|
{\cal O}^{[\Gamma]}(0; z) | p_1, s_1 \rangle.
\nonumber
\end{eqnarray}

In the $k_\perp$-independent ($k_\perp$-integrated) case, there is no a doubt that we deal with the time-reversal and parity invariance
which, together with the hermiticity, ultimately lead to the restricting conditions for the possible forms of
parametrizations through the different kinds of parton distributions (GPDs/GDAs, PDF etc) and
the corresponding Lorentz covariant tensors.
As a result, all of the parton distributions, for example GPDs, possess the
well-known symmetry properties and are the real functions \cite{Belitsky:2005qn}.
However, if the transverse momenta $k_\perp$ remain unintegrated, see the Eqn.~(\ref{gtmdf-2}),
the time-reversal invariance has been relaxed in a sense that in the Lorentz covariant parametrization
of the hadron matrix elements the time-reversal transforms do not impose any constraints
on the parametrizing functions and consequently allow the existence of the $T$-odd
parton distributions \cite{Boer:2003cm, Meissner:2008xs, Meissner:2009ww}.
Moreover, such the parton distributions as GTMDs become to be complex functions \cite{Meissner:2008xs, Meissner:2009ww}.

Indeed, for the transverse-momentum dependent functions, which take part in the parametrization
of the hadron matrix elements of the gauge-invariant operators, the time-reversal transforms lead to the
following relation:
\begin{eqnarray}
\label{T-rev}
\big\{ W^{[\Gamma]}_{[\pm]}(x, \vec{\bf k}_\perp; \overline{P}, \Delta) \big\}^*
=(-1)^\eta\, W^{[\mathds{T} \Gamma^* \mathds{T}]}_{[\mp]}(x, \widetilde{\vec{\bf k}}_\perp;
\widetilde{\overline{P}}, \widetilde{\Delta}),
\end{eqnarray}
where $\widetilde{a}=(a_0, -\vec{a})$, $\mathds{T}=i\gamma^1\gamma^3=-i\gamma_5 C$, $\eta$ defines the parity of transformations ($\eta=1$ for the odd combinations, while $\eta=2$ for the even combinations),
$[\pm]$ implies the future-pointed and the past-pointed Wilson line in the corresponding correlators.
For example, the nucleon matrix elements of the nonlocal quark operators yield to the relation (\ref{T-rev}) with $\eta=2$
for all $\Gamma$ of the Dirac $\gamma-$ basis \cite{Meissner:2008xs, Meissner:2009ww}.
It is important to emphasize that the $k_\perp$-dependent parton functions are actually the complex functions owing to
the fact that the future-pointed Wilson line is converting to the past-pointed Wilson line under the
time-reversal transforms \cite{Boer:2003cm}.

We now dwell on the nucleon GTMDs which are generated by $\gamma^+$-projection. Some of them
can be presented as
\begin{eqnarray}
\label{gtmdf-3}
&&W^{[\gamma^+]}(k; \overline{P}, \Delta) =
\\
&&
\bar u(p_2) \big[ H(x, \xi, t ;\{\vec{\bf k}_\perp\} ) \gamma^+ +
E_1(x, \xi, t ;\{\vec{\bf k}_\perp\} ) \frac{i\sigma^{+\Delta_\perp}}{2m_N} +
\nonumber\\
&&
E_2(x, \xi, t ;\{\vec{\bf k}_\perp\} ) \frac{i\sigma^{+k_\perp}}{2m_N} +
E_3(x, \xi, t ;\{\vec{\bf k}_\perp\} ) \overline{P}^+\frac{i\sigma^{k_\perp \Delta_\perp}}{2m_N^3}
\big] u(p_1)
\nonumber\\
&& + ...
\nonumber
\end{eqnarray}
where $\{\vec{\bf k}_\perp\}$ shortens the scalar product set involving
$\vec{\bf k}^2_\perp$ and $\vec{\bf k}_\perp \vec{\bm{\Delta}}_\perp$.

In Eqn.~(\ref{gtmdf-3}), the functions $H$ and $E_i$ depend on the future-pointed Wilson line and
we omit the other possible parametrizing functions which are irrelevant for our study.

Focusing on the function $E_2$, as mentioned above, the time-reversal transforms give us the
following property (see the Appendix~\ref{T-trans} for the details)
\begin{eqnarray}
\label{T-rev-1}
E_2^{[\pm]}(x, \xi, t ;\{\widetilde{\vec{\bf k}}_\perp\} ) = E_2^{[\mp]\,*}(x, \xi, t ;\{\vec{\bf k}_\perp\} ),
\end{eqnarray}
and, therefore, we have
\begin{eqnarray}
\label{T-rev-2}
&&E_2^{(\Re)\,[\pm]}(x, \xi, t ;\{\widetilde{\vec{\bf k}}_\perp\} ) = E_2^{(\Re)\,[\mp]}(x, \xi, t ;\{\vec{\bf k}_\perp\} ),
\\
&&E_2^{(\Im)\,[\pm]}(x, \xi, t ;\{\widetilde{\vec{\bf k}}_\perp\} ) = -\, E_2^{(\Im)\,[\mp]}(x, \xi, t ;\{\vec{\bf k}_\perp\} ).
\end{eqnarray}
In the $k_\perp$-integrated case, the function $E_2$ contributes to the standard twist-$2$ function $E(x,\xi,t)$,
parametrizing the collinear hadron matrix element, as \cite{Meissner:2009ww}
\begin{eqnarray}
\label{E2-E}
E(x,\xi,t) \Leftarrow \int (d^2\vec{\bf k}_\perp)
\frac{\vec{\bf k}_\perp \vec{\bm{\Delta}}_\perp}{\vec{\bm{\Delta}}^2_\perp}
E_2^{(\Re)\,[+]}(x, \xi, t ;\{\vec{\bf k}_\perp\} ).
\end{eqnarray}
Hence, we can conclude that in respect to the $\vec{\bf k}_\perp \vec{\bm{\Delta}}_\perp$-dependence,
the real part of function $E_2$ is transforming as the odd function, while the imaginary part of function $E_2$ -- as the even function,
{\it i.e.}
\begin{eqnarray}
\label{E2RI-prop}
&&E_2^{(\Re)\,[\pm]}(x, \xi, t ;\{\vec{\bf k}_\perp\} ) \sim \phi\big((\vec{\bf k}_\perp \vec{\bm{\Delta}}_\perp)^{2n+1}\big),
\\
&&E_2^{(\Im)\,[\pm]}(x, \xi, t ;\{\vec{\bf k}_\perp\} ) \sim \varphi\big((\vec{\bf k}_\perp \vec{\bm{\Delta}}_\perp)^{2n}\big).
\end{eqnarray}
On the other hand, in the forward limit defined by $\Delta=0$, the imaginary part of $E_2$ can be related
to the Sivers function $f_{1\, T}^{\perp [\pm]}$ \cite{Meissner:2009ww} as
\begin{eqnarray}
\label{Sivers}
&&E_2^{(\Im)\,[\pm]}(x, \xi ; \vec{\bf k}^2_\perp, \vec{\bf k}_\perp \vec{\bm{\Delta}}_\perp)\Big|_{\Delta=0}=
\\
&&E_2^{(\Im)\,[\pm]}(x, 0; \vec{\bf k}^2_\perp, 0)=f_{1\, T}^{\perp [\pm]}(x; \vec{\bf k}^2_\perp)
\nonumber
\end{eqnarray}

We now express the GTMD as the direct Radon transform of the $k_\perp$-dependent DD-function
$e_2(\alpha, \beta; \{\vec{\bf k}_\perp\})$, we have
\begin{eqnarray}
\label{Rel-2}
&&E_2(x, \xi, t; \{\vec{\bf k}_\perp\}) = {\cal R}_{x, \xi}  \left[ e_2(\alpha, \beta; \{\vec{\bf k}_\perp\})\right],
\\
&&
e_2(\alpha, \beta; \{\vec{\bf k}_\perp\})= e_2^{(\Re)}(\alpha, \beta; \{\vec{\bf k}_\perp\}) +
i  e_2^{(\Im)}(\alpha, \beta; \{\vec{\bf k}_\perp\}).
\nonumber
\end{eqnarray}
The $k_\perp$-dependent DD-functions can be parametrized in the similar manner as it has been done
for the usual DD-functions (cf. \cite{Belitsky:2005qn}). For instance, focusing on the
imaginary part of $e_2(\alpha, \beta; \{\vec{\bf k}_\perp\})$, we introduce the following representation:
\begin{eqnarray}
\label{Rel-2}
&&e_2^{(\Im)[\cal{C}]}(\alpha, \beta; \{\vec{\bf k}_\perp\}) =
\\
&&f_{1\, T}^{\perp [\cal{C}]}(\alpha; \vec{\bf k}^2_\perp)\,
\varphi\big((\vec{\bf k}_\perp \vec{\bm{\Delta}}_\perp)^{2n}\big)\, \pi(\alpha, \beta)
\nonumber
\end{eqnarray}
with the normalization condition
\begin{eqnarray}
\label{norm}
\int^{1-|\alpha|}_{-1+|\alpha|} d\beta\, \pi(\alpha, \beta) = 1.
\end{eqnarray}
The function $\pi(\alpha, \beta)$
describes how the longitudinal momentum transfer is shared between the partons.
The shape of $\pi$-function usually is similar to shape of a hadron distribution amplitude.
In the parametrization of $e_2^{(\Im)}$, see eqn.~(\ref{Rel-2}),
due to the nomalization conditions given by eqns.~(\ref{norm}) and (\ref{Sivers}) (the latter leads to $\varphi(0)=1$),
$f_{1\, T}^{\perp [\cal{C}]}(\alpha; \vec{\bf k}^2_\perp)$ is treated as the well-known Sivers function associated with
the Wilson line $\cal{C}$.

Concerning the inverse Radon transforms,
for the further convenience, we introduce
the shortened notation for the corresponding integration measures
($d\mu_A$ is the integration measure associated with the additional new term,
$d\mu_S$ implies the integration measure related to the standard term, see for example the eqn.~(\ref{Inv-F-t-3}))
as
\begin{eqnarray}
\label{short-measure}
&&\int d\mu_A\big\{...\big\}=
\pi \int_{-\infty}^{+\infty} (dz) \delta(z) \, \int_{-\infty}^{+\infty} d\xi
 \frac{\partial}{\partial z}\big[...\big] ,
\nonumber\\
&&
\int d\mu_S\big\{...\big\}=
\int_{-\infty}^{+\infty} (dz)\, \frac{\mathcal{P}}{z^2}\,
\int_{-\infty}^{+\infty} d\xi \, \big[...\big].
\end{eqnarray}
With these notations, we have
\begin{eqnarray}
\label{DD-inverseRtr}
&&
f_{1\, T}^{\perp [\pm]}(\alpha; \vec{\bf k}^2_\perp)\,
\varphi\big((\vec{\bf k}_\perp \vec{\bm{\Delta}}_\perp)^{2n}\big)\, \pi(\alpha, \beta)=
\nonumber\\
&&- \int d\mu_A\Big\{ E^{(\Re)[\pm]}_2(z+\alpha+\xi \beta, \xi ; \{\vec{\bf k}_\perp\}) \Big\} -
\nonumber\\
&&
\int d\mu_S\Big\{ E^{(\Im)[\pm]}_2(z+\alpha+\xi \beta, \xi ; \{\vec{\bf k}_\perp\}) \Big\}.
\end{eqnarray}
From these, we can see that {\it the inverse Radon transformations mix up the
real and imaginary parts of GTMDs thanks to the presence of the additional term that includes the
integration over the measure $d\mu_A$}. This is our principal result of the present study.
In the phenomenological application, the GTMDs can be computed within the certain model.
Hence, thanks to our main result, we can restore the Sivers function
in an alternative way, involving the new-found additional contribution, 
to be compared with the available experimental data \cite{An-Sz}.

\section{Conclusions}

We are now summarizing the main results obtained in our paper as follows.

Having considered the Radon transforms with an arbitrary real origin function $f$,
we have extended the standard representation of the inverse Radon transform
up to the physically motivated case that leads to the
certain restrictions for the angular (slope) Radon parameter.
The standard representation of the inverse Radon transform is
known in the mathematical
literature \cite{GGV} where
the integration over the full angular region ($\varphi\in [0,\,2\pi]$)
has been performed by construction.
As a result, first, both the origin function $f$ and its Radon transform have to be considered
as the complex functions and, second, the angular restrictions allow the additional term in the eqn.~(\ref{Inv-F-t-3w})exists.

By considering the mathematical example where the origin function variables $x_1$ and $x_2$ are restricted but not related
each other by any extra condition, we have
studied the influences of the restricted support of the origin function $f$ on their Fourier and
Radon transformations.
We have demonstrated that, in the most general case, the restricted support of $f$ leads to the
restrictions in the $q$-plane where the Fourier image of $f$ has been determined.
As a consequence, using the Fourier slice theorem, we have obtained in Sec.{\ref{SubSecB}}
the restrictions on the related angular Radon parameter.

Then, we have revisited the problems related to the (inverse) Radon transforms which connect the
DDs with GPDs.
Working with the support theorem for the case of origin DD-function where the variables $\alpha$ and $\beta$ are restricted and
related by the condition $\alpha + \beta < 1$ (this is the so-called spectral property),
we have shown in Sec.~\ref{B-Q-theorem} that the angular restriction for the Radon parameter, $\varphi\in [0,\,\pi]$, takes place in the
similar manner as for the considered mathematical example and leads to the existence of the additional term.
Moreover, in the frame of the obtained angular restriction,
the conditions (\ref{Con-1}) which stem from the spectral property of $\alpha$ and $\beta$ cannot ensure, generally speaking,
the physical restrictions for the GPD parameters $z$ and $\xi$
unless we have imposed $\xi \leq 1$ from the very beginning. In other words,
the spectral property of $\alpha$ and $\beta$, see the eqn.~(\ref{Theta-int-1}), cannot result directly in the
more strong restrictions for the GPD parameter $z$ and $\xi$ which lead them to their physical values.

Besides, we have investigated the symmetry properties of GPDs which are consistent with the existence of
the additional term in the eqn.~(\ref{Inv-F-t-3}).
We have discussed the evidences pointing out that this additional term can be related to the $T$-odd effects.

Finally, we have given the physical interpretation of the new additional term which we have found.
We have argued that the additional term in the eqn.~(\ref{Inv-F-t-3}) is essentially related to the $k_\perp$-dependent parton distributions.
Making the ansatz~(\ref{Rel-2}) for the transverse-momentum dependent DD-function,
we have presented the relations between the additional term and the Sivers function
given by the eqn.~(\ref{DD-inverseRtr}).


\section*{Acknowledgments}

We thank D.~M\"uller, M.V.~Polyakov, O.V.~Teryaev and the colleagues from the Theoretical Physics Division of NCBJ (Warsaw)
for useful discussions.
The work by I.V.A. was supported by the Bogoliubov-Infeld Program.
I.V.A. also thanks the Theoretical Physics Division of NCBJ (Warsaw) for warm hospitality.
L.Sz. is supported by grant No 2017/26/M/ST2/01074 of the National Science Center in
Poland and by funding from the European Union’s Horizon 2020 research and
innovation programme under grant agreement No 824093.

\appendix
\renewcommand{\theequation}{\Alph{section}.\arabic{equation}}

\section{Regularization for the Radon inversion}
\label{App1}

We dwell on the discussion of problem, as known as the ill-posedness, which appears when we need the inversion of
Radon transformations.
To begin with, we introduce the so-called dual Radon transformation defined as
\begin{eqnarray}
\label{Dual-R-1}
\mathcal{R}_\ast[g](\vec{\bf x}) &&=
\int_{-\infty}^{+\infty} d\tau \int_{0}^{2\pi} d\varphi \,g(\tau, \varphi)
\delta\left( \tau - \langle\vec{\bf n}_\varphi,\vec{\bf x}\rangle\right)
\nonumber\\
&&=\int_{0}^{2\pi} d\varphi \,g(\langle\vec{\bf n}_\varphi,\vec{\bf x}\rangle, \varphi).
\end{eqnarray}
Assuming that $g(\tau, \varphi)$ is being $\mathcal{R}[f](\tau, \varphi)$, we derive
\begin{eqnarray}
\label{Dual-R-2}
\mathcal{R}_\ast \mathcal{R}[f](\vec{\bf x}) =
\int_{0}^{2\pi} d\varphi \,\mathcal{R}[f](\langle\vec{\bf n}_\varphi,\vec{\bf x}\rangle, \varphi)
\stackrel{\text{def.}}{=} \mathcal{B}[f](\vec{\bf x})
\end{eqnarray}
where $\mathcal{B}[f](\vec{\bf x})$ denotes the back-projection for the Radon transform.

Let $\langle f, g \rangle$ be a scalar product defined on the suitable space. Therefore,
we have
\begin{eqnarray}
\label{Sc-1}
&&\langle f, \mathcal{R}_\ast\mathcal{R}[g] \rangle = \langle \mathcal{F}[f], \mathcal{F}[\mathcal{R}_\ast\mathcal{R}[g]] \rangle=
\nonumber\\
&&\int(d^2\vec{\bf q}) \overline{\mathcal{F}[f](\vec{\bf q})} \, \mathcal{F}[\mathcal{R}_\ast\mathcal{R}[g]](\vec{\bf q})=
\nonumber\\
&&
\int_{0}^{2\pi}d\varphi \int_0^{+\infty} d\lambda\, \lambda\, \overline{\mathcal{F}[f](\lambda\vec{\bf n}_\varphi)} \, \mathcal{F}[\mathcal{R}_\ast\mathcal{R}[g]](\lambda\vec{\bf n}_\varphi).
\end{eqnarray}
On the other hand, we can write this equation in the following equivalent form
\begin{eqnarray}
\label{Sc-2}
&&\langle f, \mathcal{R}_\ast\mathcal{R}[g] \rangle = \langle \mathcal{R}[f], \mathcal{R}[g] \rangle =
\nonumber\\
&&\int_{0}^{2\pi}d\varphi \int_{-\infty}^{+\infty} d\tau
\overline{\mathcal{R}[f](\tau, \varphi)}\, \mathcal{R}[g](\tau, \varphi)=
\nonumber\\
&&\int_{0}^{2\pi}d\varphi \int_0^{+\infty} d\lambda\,
\overline{\mathcal{F}[f](\lambda\vec{\bf n}_\varphi)} \, \mathcal{F}[g](\lambda\vec{\bf n}_\varphi).
\end{eqnarray}
Having compared the eqns.~(\ref{Sc-1}) and (\ref{Sc-2}), we obtain that
\begin{eqnarray}
\label{R-eig-f-1}
\lambda\, \mathcal{F}[\mathcal{R}_\ast\mathcal{R}[g]](\lambda\vec{\bf n}_\varphi) - \mathcal{F}[g](\lambda\vec{\bf n}_\varphi) =0
\end{eqnarray}
or
\begin{eqnarray}
\label{R-eig-f-2}
\mathcal{F}[\mathcal{R}_\ast\mathcal{R}[g]](\vec{\bf q}) = \frac{1}{|\vec{\bf q}|} \mathcal{F}[g](\vec{\bf q}).
\end{eqnarray}
Hence, if $\lambda\to +\infty$ then the inverse Radon transform is a singular (unbounded) one, {\it i.e.} $\mathcal{R}^{-1}\to +\infty$.
As a result, the inverse Radon transformation needs to be regularized.

Notice that
\begin{eqnarray}
\label{Reg-1}
\mathcal{F}^{-1}\Big[\frac{1}{|\vec{\bf q}|}\Big](\vec{\bf x})= \int_{0}^{2\pi}d\varphi
\int_0^{+\infty} d\lambda\, e^{-i\lambda\langle\vec{\bf n}_\varphi,\vec{\bf x}\rangle},
\end{eqnarray}
where we regularize the integration over $d\lambda$ as
\begin{eqnarray}
\label{Reg-2}
\lim_{\varepsilon\to 0}\int_0^{+\infty} d\lambda\,
e^{-i\lambda \left(\langle\vec{\bf n}_\varphi,\vec{\bf x}\rangle-i\varepsilon\right)}=
\lim_{\varepsilon\to 0} \,\frac{-i}{\langle\vec{\bf n}_\varphi,\vec{\bf x}\rangle-i\varepsilon}.
\end{eqnarray}
Therefore, we have
\begin{eqnarray}
\label{Reg-3}
\mathcal{F}^{-1}\Big[\frac{1}{|\vec{\bf q}|}\Big]_{reg}(\vec{\bf x}) = \mathcal{R}_\ast\Big[\frac{-i}{\tau - i\varepsilon} \Big](\vec{\bf x}).
\end{eqnarray}
One can immediately see that eqn.~(\ref{Reg-3}) leads to the necessity of regularization which we have performed in eqn.~(\ref{int-lam}).
To conclude this section, we have demonstrated that the ill-posedness regarding the inverse Radon transformation
can be treated with a help of suitable regularization in the integration over $\lambda$ as in eqn.~(\ref{int-lam}).

\section{The time-reversal property}
\label{T-trans}

For the pedagogical reason, we are presenting the details how to extract the given property
of function $E_2$ under the time-reversal transform.

To begin with, we write the time-reversal transformation for the non-forward matrix element of $\gamma^+$-projection , we have
\begin{eqnarray}
\label{T-corr}
&&\langle \widetilde{p}_2, \widetilde{s}_2 | \bar\psi(0) \widetilde{\gamma}^+\psi(\widetilde{z}^-,\widetilde{{\bf z}}_\perp
|\widetilde{p}_1, \widetilde{s}_1 \rangle =
\nonumber\\
&&
\Big\{ \langle p_2, s_2 | \bar\psi(0) \gamma^+\psi(z^-,{\bf z}_\perp
|p_1, s_1 \rangle\Big\}^*,
\end{eqnarray}
where
\begin{eqnarray}
\label{tilde-gamma-plus}
\widetilde{\gamma}^+ = \mathds{T}\, \gamma^+\, \mathds{T}= \gamma^-.
\end{eqnarray}

Let us write the {\it l.h.s.} of Eqn.~(\ref{T-corr}) with a help of parametrization through the function $E_2$. It reads
\begin{eqnarray}
\label{T-corr-lhs-1}
&&\langle \widetilde{p}_2, \widetilde{s}_2 | \bar\psi(0) \gamma^-\psi(\widetilde{z}^-,\widetilde{{\bf z}}_\perp
|\widetilde{p}_1, \widetilde{s}_1 \rangle =
\nonumber\\
&&
\int (d\widetilde{k}^+ d^2\widetilde{\vec{\bf k}}_\perp) e^{-i\widetilde{K}^+ \widetilde{z}^-
+i \widetilde{\vec{\bf k}}_\perp\widetilde{\vec{\bf z}}_\perp}
E_2\big(\widetilde{k}^+, \widetilde{\Delta}^+; \{\widetilde{\vec{\bf k}}_\perp\}\big)\times
\nonumber\\
&&\frac{1}{2m_N}\,
\bar u(\widetilde{p}_2, \widetilde{s}_2) i \sigma^{-\,\widetilde{\vec{\bf k}}_\perp} u(\widetilde{p}_1, \widetilde{s}_1).
\end{eqnarray}
Then, using $\mathds{T}^2=1$, we get
\begin{eqnarray}
\label{T-2}
&&\bar u(\widetilde{p}_2, \widetilde{s}_2) \mathds{T} \,
\Big\{ \mathds{T}i \sigma^{-\,\widetilde{\vec{\bf k}}_\perp} \mathds{T}\Big\}
\, \mathds{T}u(\widetilde{p}_1, \widetilde{s}_1)=
\nonumber\\
&&\Big\{
\bar u(p_2, s_2)\, i \sigma^{+\,\vec{\bf k}_\perp} \,u(p_1, s_1)
\Big\}^*
\end{eqnarray}
which coincides with the spinor structure of the {\it r.h.s.} of Eqn.~(\ref{T-corr}).
As well-known, the time-reversal transforms convert the dominant plus light-cone direction
to the dominant minus light-cone direction. Due to the fact that the variables $x$ and $\xi$ are
invariant on the light-cone,
\begin{eqnarray}
x=\frac{k^+}{\overline{P}^+}=\frac{k_+}{\overline{P}_+},\quad
-2\xi=\frac{\Delta^+}{\overline{P}^+}=\frac{\Delta_+}{\overline{P}_+},
\end{eqnarray}
we therefore obtain the time-reversal property as
\begin{eqnarray}
\label{T-rev-1A}
E_2^{[\pm]}(x, \xi, t ;\{\widetilde{\vec{\bf k}}_\perp\} ) = E_2^{[\mp]\,*}(x, \xi, t ;\{\vec{\bf k}_\perp\} ).
\end{eqnarray}
In the same manner we can derive the analogous properties for the other parametrizing functions.


\begin{thebibliography}{99}
\vspace{1\baselineskip}

\bibitem{DM} D.M{\"u}ller et al., Fortschr. Phys. {\bf 42}, 101 (1994).

\bibitem{Ji:1996nm}
  X.~D.~Ji,
  Phys.\ Rev.\ D {\bf 55}, 7114 (1997)

\bibitem{Radyushkin:1997ki}
  A.~V.~Radyushkin,
  Phys.\ Rev.\ D {\bf 56}, 5524 (1997)

\bibitem{Ball:1996tb}
  P.~Ball and V.~M.~Braun,
  Phys.\ Rev.\ D {\bf 54}, 2182 (1996)

\bibitem{Braun:1999te}
  V.~M.~Braun, S.~E.~Derkachov, G.~P.~Korchemsky and A.~N.~Manashov,
  Nucl.\ Phys.\ B {\bf 553}, 355 (1999)

\bibitem{Diehl:2003ny}
  M.~Diehl,
  Phys.\ Rept.\  {\bf 388}, 41 (2003)

\bibitem{Belitsky:2005qn}
  A.~V.~Belitsky and A.~V.~Radyushkin,
  Phys.\ Rept.\  {\bf 418}, 1 (2005)

\bibitem{Teryaev:2001qm}
  O.~V.~Teryaev,
  Phys.\ Lett.\ B {\bf 510}, 125 (2001)

\bibitem{Muller:2017wms}
  D.~Müller,
  ``Double distributions and generalized parton distributions from the parton number conserved light front wave function overlap representation,''
  arXiv:1711.09932 [hep-ph].

\bibitem{HMud1}
  N.~Chouika, C.~Mezrag, H.~Moutarde and J.~Rodríguez-Quintero,
  Eur.\ Phys.\ J.\ C {\bf 77},  906 (2017) .

\bibitem{HMud2}
  N.~Chouika, C.~Mezrag, H.~Moutarde and J.~Rodríguez-Quintero,
  EPJ Web Conf.\  {\bf 137}, 05020  (2017).

\bibitem{Ch-Th}
  N.~Chouika,
  ``Generalized Parton Distributions and their covariant extension: towards nucleon tomography,''
  2018SACLS259, tel-01925746, PhD thesis.

\bibitem{HM-Th}
  H.~Moutarde,
  ``Nucleon Reverse Engineering: Structuring hadrons with colored degrees of freedom,''
Habilitation thesis, 243 p (2015).


\bibitem{Deans}
S.R.~Deans, ``The Radon Transform and Some of Its Applications,'' Wiley 299 p (1983).

\bibitem{GGV}
I.M.~Gelfand, M.I.~Graev, N.Ya.~Vilenkin,
``Generalized Functions, Volume 5: Integral Geometry and Representation Theory,''
AMS Chelsea Publishing: An Imprint of the American Mathematical Society, 449 pp (1966).

\bibitem{BQ-th}
J.~Boman and E.T.~ Quinto,
Duke Math.\ J. {\bf 55}, 943 (1987).


\bibitem{Gabdrakhmanov:2019bqm}
  I.~R.~Gabdrakhmanov, D.~Müller and O.~V.~Teryaev,
  ``Inverse Radon transform at work,''
  arXiv:1906.01458 [hep-ph].

\bibitem{Boer:2003cm}
  D.~Boer, P.~J.~Mulders and F.~Pijlman,
  Nucl.\ Phys.\ B {\bf 667}, 201 (2003)

\bibitem{Meissner:2008xs}
  S.~Meissner, A.~Metz and M.~Schlegel,
  ``Generalized transverse momentum dependent parton distributions of the nucleon,''
 arXiv:0807.1154 [hep-ph].

\bibitem{Meissner:2009ww}
  S.~Meissner, A.~Metz and M.~Schlegel,
  JHEP {\bf 0908}, 056 (2009)

\bibitem{An-Sz}
  I.V.~Anikin and L.~Szymanowski, work in progress.

\end{thebibliography}
\end{document}